\documentclass[aps,preprint]{revtex4}%
\usepackage{amsfonts}
\usepackage{amsmath}
\usepackage{amssymb}
\usepackage{graphicx}%
\providecommand{\U}[1]{\protect\rule{.1in}{.1in}}

\begin{document}
\preprint{ }
\title{Is the Universe Flat?}
\author{Sel\c{c}uk \c{S}. Bayin}
\affiliation{Middle East Technical University, Institute of applied Mathematics, Ankara,
06800, Turkey}
\keywords{Cosmology; Geometry of the universe; Gravitational temperature; General relativity}
\pacs{04.20.-q; 04.50.Kd; 04.62.+v }

\begin{abstract}
Geometry of the universe has always intrigued mathematicians and cosmologists.
Recent results from the Wilkinson Microwave Anisotropy Project (WMAP) indicate
that the visible universe is incredibly flat. This apparent flatness could be
due to the fact that only a small part of the universe is visible, thus
indicating that the geometry of the universe is still an open and an
interesting problem in cosmology.\textbf{ }Assuming a profound connection
between Friedmann, Robertson, Walker (FRW) geometries and thermodynamics, we
construct a parameter free exploratory model that allows us to predict the
geometry of the universe by thermodynamic arguments. The key parameters in
this model are the concept of \textit{global equation of state} and the
concept of \textit{gravitational temperature}. By comparing the equal time
expansion of the Green function for the massless conformal scalar field in
background FRW geometry\textbf{ }with the thermal Green function in Minkowski
space-time, we define the gravitational temperature. We also give the protocol
for determining the global equation of state for a given local equation of
state. Using a local equation of state that covers a wide range of physically
acceptable cases, $P=\alpha\rho,\alpha>0,$ and within the context of FRW
thermodynamics, we predict that the geometry of the visible universe is
Lobachevskian (open), albeit being very close to flat. This is consistent with
the WMAP data, which indicates that the universe may deviate from flatness by
as much as 1\%. We also discuss the self-consistency of this suggestive model
along with its possible connections with nonextensive thermodynamics and black
hole thermodynamics.

\end{abstract}
\maketitle
\date{}

\section{Introduction}

A profound connection between geometry and thermodynamics was first hinted in
1970s by\ the works of Bekenstein [1] and Hawking [2] with the black hole
thermodynamics. Later, the discovery of black hole radiation by Hawking
established the physical link between the area of the horizon with entropy and
the surface gravity of the black hole with temperature. However, despite the
success of black hole thermodynamics, the relation between general theory of
relativity and thermodynamics remains to be established for more general metrics.

In 1995, assuming \textbf{extensivity} of entropy, Jacobson [3] obtained
Einstein's field equations from the proportionality of entropy and the area of
the horizon, and the relation%
\begin{equation}
\delta Q=TdS, \label{1}%
\end{equation}
where $Q$ is heat, $S$ is entropy, and $T$ is the temperature. Jacobson's
treatment hinges on the fact that causal horizons hide information, hence
should be associated with entropy, where heat is defined as the energy that
flows through the causal horizon that continues to interact with the outside
gravitationally. Recently, assuming that the holographic principle holds, and
building on the works of Jacobson [3] and Padmanabhan [4], Verlinde [5]
proposed a framework for gravity as an entropic force via the assumption that
holographic principal holds. On the nonextensive thermodynamics side, recently
Tsallis and Cirto [6] argued that gravitating systems should be associated
with an appropriate \textbf{nonextensive} generalization of the additive
expression
\begin{equation}
S=k_{B}\ln W. \label{2}%
\end{equation}

The fact that horizons hide information motivates their association with
entropy, but due to the universality of gravity and the principle of
equivalence, we can also expect a more general relation between gravity and
entropy, independent of the existence of a horizon. This follows from the fact
that all one can say about the source of a gravitational field is its total
energy-momentum distribution. Since there are many different ways to generate
the same energy-momentum distribution, the gravitational field itself embodies
a certain amount of information. Even though classically the number of
possible sources that could produce a given gravitational field is infinity,
we expect quantum mechanics to give a finite number.

On the thermodynamic side of this captivating connection between gravitation
and thermodynamics, we revisit the suggestive model introduced in 1986 as the
Friedmann thermodynamics [7-10] and give new derivations of its key components
like the \textit{gravitational\ temperature} and the \textit{global equation
of state}. By comparing the equal time expansion of the Green function of the
massless conformal scalar field in background FRW geometry with the thermal
Green function in Minkowski space-time, we give a new derivation of\textit{
}the\textit{ }gravitational temperature. Using this temperature within the
context of Friedmann thermodynamics, which we also reestablish, we argue that
the geometry of the visible universe is Lobachevskian. In this regard, the
lack of any measurable curvature in current observations means that the
universe\ must be much larger than what is visible to us. We also discuss
limitations and the logical consistency of this model, along with its
connections with nonextensive thermodynamics.

\section{Geometry and Einstein Cosmology in a Nutshell}

The large scale homogeneity and isotropy of the universe suggest that the
universe at large can be described by one of the FRW models, which are
represented by the line element%
\begin{align}
ds^{2}  &  =g_{\mu\nu}dx^{\mu}dx^{\nu},\text{ }\mu,\nu=0,1,2,3,\nonumber\\
&  =dt^{2}-e^{g(t)}\left[  \frac{1}{\left(  1-k\frac{r^{2}}{R_{0}^{2}}\right)
}dr^{2}+r^{2}d^{2}\Omega\right]  ,\text{ }k=0,\pm1, \label{3}%
\end{align}
\ where $g_{\mu\nu}$ is the FRW metric that gives a complete description of
the \textit{local} space-time geometry. In FRW models the spatial universe is
described by the constant time slices of the line element. For $k=0$ the
universe is flat and the geometry is Euclidean, for $k=1$ the geometry is
Gaussian and the universe is the three dimensional surface of a four
dimensional hypersphere with time dependent radius $R(t)=R_{0}e^{g(t)/2}$, for
$k=-1$ the geometry is Lobachevskian, where the universe can be considered as
the surface of a hypersphere with imaginary radius $R(t)=iR_{0}e^{g(t)/2}$.
The FRW models are in general time dependent, where \textbf{ }$R_{0}%
e^{g(t)/2}$ is the scale factor. We will take the initial conditions so that
$R_{0}$ is the present radius of the universe. One of the trademarks of a
given geometry is the sum of the interior angles of triangles, which for the
Euclidean geometry is always equal to $\pi$ and greater than $\pi$ for the
Gaussian, and less than $\pi$ and for the Lobachevskian geometries.

Even though geometry is a directly observable feature of the universe through
angle, distance, area, etc. measurements, cosmic scales make this rather
impractical. On the other hand, Einstein's field equations:%

\begin{equation}
G^{\mu\nu}=-8\pi T^{\mu\nu},\text{ }c=G=1, \label{4}%
\end{equation}
which relate the space-time geometry to the matter content of the universe,
offer a much more practical alternative. The left hand side of the Einstein's
field equations, $G^{\mu\nu},$ is called the Einstein tensor, which is a
purely geometric quantity entirely composed of the metric tensor and its first
two derivatives. The right hand side, $T^{\mu\nu},$ is the energy-momentum
tensor, which describes the matter content of the universe. In this work, we
treat the cosmological constant as a part of the energy-momentum tensor, hence
do not write it explicitly.

For perfect fluids the energy-momentum tensor is given as
\begin{equation}
T^{\mu\nu}=(P_{0}+\rho_{0})U^{\mu}U^{\nu}-P_{0}g^{\mu\nu}, \label{5}%
\end{equation}
where in comoving\ coordinates we set $U^{1}=U^{2}=U^{3}=0.$ With the FRW
metric [Eq. (\ref{3})], the field equations reduce to two equations in three
unknowns, $P_{0},\rho_{0},R,$ as%
\begin{align}
\overset{..}{R}(t)  &  =-4\pi(\rho_{0}+3P_{0})\frac{R}{3},\label{6}\\
\left(  \frac{3}{8\pi R^{2}}\right)  k  &  =\left(  \rho_{0}-\frac{3H(t)^{2}%
}{8\pi}\right)  . \label{7}%
\end{align}
In these equations, $P_{0}$ is\ the pressure, $\rho_{0}$ is the mass density
and $H(t)=\overset{.}{R}/R$ is the Hubble parameter of the universe.
Supplemented with the necessary information about the matter content of the
universe, that is, an equation of state, $P_{0}=P_{0}(\rho_{0}),$ the above
system of equations can be solved \ for a given geometry, that is, for a given
value of $k,$ to yield the solution as $\left\{  P_{0}(t),\rho_{0}%
(t),R(t)\right\}  $.

Except for exotic forms of matter $(\rho_{0}+3P_{0})$ is positive, hence the
first field equation [Eq. (\ref{6})] indicates that the universe is
decelerating. Since $k$ is a constant, we can set the values of the quantities
in Equation (\ref{7}) to their current values, thereby reducing the problem of
determining the geometry of the universe to observing the difference between
the current density and the critical density, that is, the sign of
\begin{equation}
k=\frac{\rho_{0}(t_{now})-\rho_{c}}{\left\vert \rho_{0}(t_{now)}-\rho
_{c}\right\vert }, \label{8}%
\end{equation}
where, $\rho_{c}=\frac{3H_{0}^{2}}{8\pi}$ is called the critical density and
$H_{0}$ is the Hubble constant $H(t_{now})$. Now the geometry of the universe
is%
\begin{equation}
k=\left\{
\begin{tabular}
[c]{lllll}%
$1$ & for & $\rho_{0}(t_{now})>\rho_{c}$ & ; & Gaussian\\
$0$ & for & $\rho_{0}(t_{now})=\rho_{c}$ & ; & Euclidean\\
$-1$ & for & $\rho_{0}(t_{now})<\rho_{c}$ & ; & Lobachevskian
\end{tabular}
\ \ \ \right.  . \label{9}%
\end{equation}

From the redshift measurements of galaxies, the current value of the Hubble
parameter indicates that $\rho_{c}$ is around $10^{-29}gm/cc$. On the other
hand, the dynamic mass measurements indicate that the amount of luminous
matter plus the dark matter in the universe only adds up to around $30$ $\%$
of the critical density, which from Equation (\ref{9}) indicates that we live
in a Lobachevskian universe. Just when all such data pointed to a
Lobachevskian universe, the WMAP satellite measurements of the anisotropy of
the cosmic microwave background radiation surprised cosmologists by yielding a
$k\ $value very close to zero. This also meant that the current density of the
universe is very close to the critical density, thus indicating that almost
$70\%$ of the matter content of the universe is yet to be accounted for.

Another surprise came with the discovery of the acceleration of the universe,
which allowed us to interpret this missing $70\%$ in terms of some exotic form
of matter called the \textit{dark energy}. Even though the physics of dark
energy is still unknown, we can say that it has positive mass and responds to
gravity with repulsion. This naturally demands a radical change in our
understanding of the contents of the universe. Considering that a flat
universe demands a precisely tuned current density to the critical density,
which is a tall order for any observation, it is only fair to say that the
geometry of the universe will still continue to captivate cosmologists for
years to come. In fact, the recent data from the WMAP indicate that the total
density differs from the critical density by as much as 1\% or so [11].

Since FRW models with different $k$ values correspond to different spatial
distributions of galaxies, we address this problem by investigating the
connection between geometry and thermodynamics. In thermodynamics, a
particular crystal structure, that is, a particular spatial distribution of
atoms, becomes the stable form of matter with respect to the entropy criteria.
In this regard, we think that it may be possible to predict the geometry of
the universe via some \textit{thermodynamic} arguments. A profound connection
between geometry and thermodynamics was first hinted by the black hole
thermodynamics [1,2,12]. We now further the suggestive (toy) model we
introduced in 1986 as the \textit{Friedmann thermodynamics }and clarify some
of its critical concepts [7-10]\textbf{, }which has some definite predictions
about the geometry of the universe. In Section III, we introduce the concept
of \textit{global equation of state}, which plays a central role in this
model. In Section IV, we give a derivation of the \textit{gravitational
temperature} for the FRW geometries. In Section V, we introduce the
\textit{local }equation of state concept and discuss the assumptions involved.
In Section VI, for a given local equation of state, we discuss how to find the
global equation of state and then how to determine the geometry of the
universe. In Section VII, using a local equation of state that covers a wide
range of physically acceptable cases, we argue that with respect to the
Friedmann thermodynamics, the geometry of the universe is Lobachevskian.
Finally, in Section VIII, we discuss our results.

\section{The Global Equation of State}

Einstein once said that the left hand side of the field equations is as solid
as the rock of Gibraltar but the right hand side is like a house built from a
deck of cards [Eq. (\ref{4})].\ The left hand side of the Einstein's field
equations, $G^{\mu\nu}$, is a unique divergenceless tensor, entirely composed
of the metric tensor and its first two derivatives. On the other hand, the
right hand side is the energy-momentum tensor, $T^{\mu\nu}$, which represents
the matter content of the universe. The problem is due to the fact that matter
not only affects geometry but also gets affected by it, hence $T^{\mu\nu}$ is
also a function of the unknown, that is, the metric tensor. Unlike the special
theory of relativity, where one can obtain the energy-momentum tensor in a
moving frame from the rest frame expression by a Lorentz transformation, it is
not possible to generate the curved space-time energy-momentum tensor,
$T^{\mu\nu}(g^{\mu\nu}),$ from its flat (Minkowski) space-time expression by a
general coordinate transformation. In flat space-time\textbf{,} that is, in a
free fall frame, the perfect fluid energy-momentum tensor is written as
\begin{equation}
T_{0}^{\mu\nu}=(P_{0}+\rho_{0})u^{\mu}u^{\nu}-P_{0}\eta^{\mu\nu}, \label{10}%
\end{equation}
where $P_{0}$ and $\rho_{0}$ are the pressure and the density distributions,
respectively. In principle, there are infinitely many expressions that one
could write for the curved space-time energy momentum tensor. Of course, all
of them reducing to Equation (\ref{10}) in the limit $g^{\mu\nu}%
\rightarrow\eta^{\mu\nu}.$ For the energy-momentum tensor in curved
space-time, the common practice is to replace the Minkowski metric in
$T_{0}^{\mu\nu}$ with the general curved space-time metric, $g^{\mu\nu},$ and
express $P_{0}$ and $\rho_{0}$ in curved space-time coordinates to write \
\begin{equation}
T^{\mu\nu}=(P_{0}+\rho_{0})u^{\mu}u^{\nu}-P_{0}g^{\mu\nu}. \label{11}%
\end{equation}
Since $P_{0}$ and $\rho_{0}$ are defined \textit{locally}, their values at a
given point in curved space-time coordinates do not change, thus the equation
of state in curved space-time is still taken as the flat space-time equation
of state, that is, as $P_{0}=P_{0}(\rho_{0}).$

Friedmann thermodynamics is based on the observation that different crystal
structures of matter correspond to different spatial distributions of atoms
over lattices\textbf{ }with discrete symmetry properties. At a given
temperature, what defines the stable phase is the crystal structure that has
the lowest Gibbs energy. Analogously, the FRW models correspond to different
spatial distributions of galaxies, where the space over which the galaxies are
distributed is\ curved over a hyperspace with continuous but nevertheless with
distinct symmetry properties called the Bianchi symmetries.

Phase transitions are due to collective behavior of matter, hence they are
governed by the entropy criteria rather than the energy. For example, tin has
two allotropic forms at normal pressure as white and gray tin, where white tin
exists in stable form at 298 K, a form that should be unstable according to
the energy criteria [13]. \ Similarly, we can expect to determine the geometry
of the universe via thermodynamic arguments, but the conventional
energy-momentum tensor [Eq. (\ref{11})] with a given flat space-time equation
of state, $P_{0}=P_{0}(\rho_{0}),$ yields isentropic (or isothermal) models
that all have the same Gibbs energy. In other words, with the conventional
approach, all FRW models have the same Gibbs energy, $\rho_{0}=\rho_{0}%
(P_{0})$, which leaves $k$ as a parameter that can only be determined by observation.

In 1902 Gibbs [14] pointed out that systems with long range interactions like
gravity, are in principle beyond the scope of standard (Boltzmann-Gibbs)
statistical mechanics. In line with Gibbs, we now argue that thermodynamic
systems that are large enough for the effects of curvature to be important,
are endowed with a \textit{gravitational temperature} and a
\textit{gravitational entropy,} and define the curved space-time
energy-momentum tensor as%
\begin{equation}
T^{\mu\nu}=(P+\rho)u^{\mu}u^{\nu}-Pg^{\mu\nu}, \label{12}%
\end{equation}
where $P(\rho,T)$ and $\rho(P,T)$ are the \textit{global pressure }and
\textit{density} distributions, respectively, and $T$ is the
\textit{gravitational temperature}. In this approach, the vacuum field
equations remain intact, but the right hand side of the field equations is
modified profoundly so that the effect of curvature on the energy-momentum
tensor is included not only explicitly through the replacement $\eta^{\mu\nu
}\rightarrow g^{\mu\nu},$ but also implicitly through the global equation of
state $P=P(\rho,T)$. As we will show, compared to other modifications of the
Einstein's field equations, this is a rather intricate change that allows us
to predict the geometry of the universe. Note that the new energy-momentum
tensor, as it should, in the Minkowski limit, $g^{\mu\nu}\rightarrow\eta
^{\mu\nu},$ reduces to $T_{0}^{\mu\nu},$ where $P\rightarrow P_{0}$ and
$\rho\rightarrow\rho_{0}.$

\section{Gravitational Temperature}

To define a \textit{gravitational} \textit{temperature} for the FRW
geometries, we use the massless conformal scalar field in curved background
FRW geometry. The corresponding vacuum Wightman functions, $G_{(k)}%
^{(+)}(x,x^{\prime})=\left\langle 0\left\vert \phi(x)\phi(x^{\prime
})\right\vert 0\right\rangle ,$ in static FRW universes are given as [15]%
\begin{align}
G_{(1)}^{(+)}(x,x^{\prime})  &  =\frac{1}{8\pi^{2}R^{2}\left[  \cos(\Delta
\eta-i\epsilon)-\cos(\Delta\chi)\right]  },\text{ }\sin\chi=\frac{r}{R},\text{
}\chi\in\lbrack0,\pi],\label{13}\\
G_{(0)}^{(+)}(x,x^{\prime})  &  =-\frac{1}{4\pi^{2}}\frac{1}{\Delta
t^{2}-\Delta r^{2}},\label{14}\\
G_{(-1)}^{(+)}(x,x^{\prime})  &  =\frac{\Delta\chi}{4\pi^{2}R^{2}\sinh
(\Delta\chi)\left[  \Delta\chi^{2}-(\Delta\eta-i\epsilon)^{2}\right]  },\text{
}\sinh\chi=\frac{r}{R},\text{ }\chi\in\lbrack0,\infty], \label{15}%
\end{align}
where $R$ is the constant scale factor and $\eta$ is the conformal time.

To find the gravitational temperature, we use the thermal Green function of
the massless conformal scalar field obeying Maxwell-Boltzmann statistics,
which can be written as an infinite sum over imaginary time as [15]%
\begin{equation}
G_{\beta}^{(1)}(x,x^{\prime})=\sum_{n=-\infty}^{\infty}G^{(1)}(t+in\beta
,\overrightarrow{r},t^{\prime},\overrightarrow{r}^{\prime}),\text{ }%
\beta=\frac{1}{T},\text{ }k_{B}=\hslash=1, \label{16}%
\end{equation}
where $G^{(1)}=\left\langle 0\left\vert \left\{  \phi(x),\phi(x^{\prime
})\right\}  \right\vert 0\right\rangle $ is the zero temperature Green function.

Using the zero temperature Green function of the massless conformal scalar
field in Minkowski space-time:
\begin{equation}
^{M}G^{(1)}(x,x^{\prime})=-\left(  \frac{2}{4\pi^{2}}\right)  \frac{1}{\Delta
t^{2}-\Delta\rho^{2}},\text{ }\Delta t=t-t^{\prime},\text{ }\Delta
\rho=r-r^{\prime}, \label{17}%
\end{equation}
we write the corresponding thermal Green function as%
\begin{equation}
^{M}G_{\beta}^{(1)}(x,x^{\prime})=-\frac{1}{2\pi^{2}}\sum_{n=-\infty}^{\infty
}\frac{1}{\left(  \Delta t+in\beta\right)  ^{2}-\Delta\rho^{2}}. \label{18}%
\end{equation}
Following Mamaev and Trunov [16], we now write the equal time, $\Delta t=0,$
expansion of $^{M}G_{\beta}^{(1)}(x,x^{\prime})$ in powers of $\Delta\rho$ as%

\begin{equation}
^{M}G_{\beta}^{(1)}(x,x^{\prime})\simeq\frac{1}{2\pi^{2}\Delta\rho^{2}}%
+\frac{T^{2}}{6}+0(\Delta\rho^{2}). \label{19}%
\end{equation}
We also write the equal time expansion of the Green function $G_{(k)}%
^{(1)}=2\operatorname{Re}G_{(k)}^{(+)},$ for the massless conformal scalar
field in static FRW universe with $k=1$ as%
\begin{equation}
G_{(1)}^{(1)}\simeq\frac{1}{2\pi^{2}\Delta\rho^{2}}+\ \frac{1}{3(8\pi
^{2})R^{2}}+0\left(  \left(  \Delta\rho/R\right)  ^{2}\right)  ,\text{ }%
\Delta\rho=R\Delta\chi, \label{20}%
\end{equation}
which when compared with Equation (\ref{19}) yields the\textit{ gravitational
temperature:}
\begin{equation}
T=\frac{1}{2\pi R},\text{ }k=1. \label{21}%
\end{equation}
This is in agreement with the \textit{effective gravitational temperature}
deduced from the Casimir effect calculations [8].

Similarly, for $k=-1,$ we write the equal time expansion for the massless
conformal scalar field as%
\begin{equation}
G_{(-1)}^{(1)}\simeq\frac{1}{2\pi^{2}\Delta\rho^{2}}-\frac{1}{12\pi^{2}R^{2}%
}+0\left(  \left(  \Delta\rho/R\right)  ^{2}\right)  ,\text{ }\Delta
\rho=R\Delta\chi, \label{22}%
\end{equation}
which when compared with Equation (\ref{19}), yields an imaginary
gravitational temperature
\begin{equation}
T=i\frac{\sqrt{2}}{2\pi R},\text{ }k=-1. \label{23}%
\end{equation}
Note that in this case, the Casimir effect calculations give only the real
part of the effective gravitational temperature, which is zero [8].

For sufficiently slowly expanding universes, where particle creation can be
ignored, the gravitational temperature can be generalized simply by replacing
$R$ with its time dependent expression, $R(t)=R_{0}e^{g(t)/2}$. Similarly,
granted that the time dependence is sufficiently slow, the effective
gravitational temperature for the time dependent flat, $k=0,$ FRW models is
defined as zero\textbf{. }

\subsection{Connection with the Unruh Temperature}

To gain a better understanding of what this \textit{gravitational temperature}
means, we apply this method to an accelerated particle detector coupled to a
massless conformal scalar field, where the Green function is given as [15]
\begin{equation}
G^{(+)}(x,x^{\prime})=-\frac{1}{16\pi^{2}\alpha^{2}}\sinh^{-2}\left(
\frac{\tau-\tau^{\prime}}{2\alpha}\right)  ,\text{ }\alpha=\text{const.}
\label{24}%
\end{equation}
The proper time $\tau$ is related to $t$ as%
\begin{equation}
t=\alpha\sinh(\tau/\alpha), \label{25}%
\end{equation}
where the detector moves along a hyperbolic trajectory with the proper
acceleration $\omega=1/\alpha.$ Using $G^{(1)}=2\operatorname{Re}G^{(+)},$ we
can write the equal time expansion as%
\begin{equation}
G^{(1)}=-\frac{1}{2\pi^{2}\Delta\tau^{2}}+\frac{\omega^{2}}{24\pi^{2}%
}+O\left(  \Delta\tau^{2}\right)  . \label{26}%
\end{equation}
Comparing this with the expansion of the thermal Green function in Equation
(\ref{19}) with $t=\tau$:
\begin{equation}
^{M}G_{\beta}^{(1)}(x,x^{\prime})\simeq-\frac{1}{2\pi^{2}}\left[  \frac
{1}{\Delta\tau^{2}-\Delta\rho^{2}}-\frac{\pi^{2}T^{2}}{3}\right]
+0(\Delta\tau^{2}-\Delta\rho^{2}), \label{27}%
\end{equation}
and with $\Delta\rho=0,$ and as $\Delta\tau\rightarrow0:$%
\begin{equation}
^{M}G_{\beta}^{(1)}(x,x^{\prime})\simeq-\frac{1}{2\pi^{2}}\left[  \frac
{1}{\Delta\tau^{2}}-\frac{\pi^{2}T^{2}}{3}\right]  +0(\Delta\tau^{2}),
\label{28}%
\end{equation}
we obtain the Unruh temperature [15]
\begin{equation}
T=\frac{\omega}{2\pi}. \label{29}%
\end{equation}

Note that for the uniformly accelerated detector [Eq. (\ref{24})], using the
identity
\begin{equation}
\csc^{2}\pi x=\pi^{-2}\sum_{k=-\infty}^{\infty}(x-k)^{-2} \label{30}%
\end{equation}
and the relation $G^{(1)}=2\operatorname{Re}G^{(+)},$ we can also write
$G^{(1)}(\Delta\tau)$ as
\begin{equation}
G^{(1)}(\Delta\tau)=-(2\pi^{2})^{-1}\sum_{k=-\infty}^{\infty}(\Delta\tau+2\pi
i\alpha k)^{-2}. \label{31}%
\end{equation}
Comparing this with Equation (\ref{18}) and with the substitutions $t=\tau$
and $\Delta\rho=0$:
\begin{equation}
^{M}G_{\beta}^{(1)}(x,x^{\prime})=-\frac{1}{2\pi^{2}}\sum_{n=-\infty}^{\infty
}\frac{1}{\left(  \Delta\tau+in\beta\right)  ^{2}}, \label{32}%
\end{equation}
we see that the two Green functions are actually identical to all orders for
the Unruh temperature $T=1/2\pi\alpha.$

In other words, for the massless conformal scalar field, the vacuum Green
function for the uniformly accelerating detector is identical to the thermal
Green function of an inertial detector at the Unruh temperature. Since the
Green functions for the FRW models, $G_{(1)}^{(1)},$ $G_{(0)}^{(1)},$
$G_{(-1)}^{(1)},$ differ from $^{M}G_{\beta}^{(1)}$ to second and higher
orders in $\Delta\rho/R,$ the \textit{gravitational temperature} we deduce is
only an \textit{effective temperature }in terms of the Maxwell-Boltzmann
statistics. Potential implications of this will be discussed in Section VIII.

\section{The Local Equation of State}

In Friedmann thermodynamics, the flat space-time equation of state,
$P_{0}=P_{0}(\rho_{0}),$ is also the \textit{local equation of state}. That
is, the equation of state that one would observe when the size of the
thermodynamic system is sufficiently small compared to the current radius of
the universe:%
\begin{equation}
r\ll R_{0}, \label{33}%
\end{equation}
where $r$ is any point within the local system. Now, the FRW line element [Eq.
(\ref{3})] can be expanded as
\begin{equation}
ds^{2}\simeq dt^{2}-e^{g(t)}\left[  \left(  1+k\frac{r^{2}}{R_{0}^{2}}%
+\cdots\right)  dr^{2}+r^{2}d^{2}\Omega\right]  ,\text{ }k=\pm1, \label{34}%
\end{equation}
and the field equations become%
\begin{align}
8\pi\rho &  \simeq3\frac{\overset{.}{R}^{2}}{R^{2}}\left(  1-k\frac{r^{2}%
}{R_{0}^{2}}+\cdots\right)  +\frac{3k}{R^{2}}\left(  1-\frac{5}{3}k\frac
{r^{2}}{R_{0}^{2}}+\cdots\right)  ,\label{35}\\
8\pi P  &  \simeq-2\frac{\overset{..}{R}}{R}-\frac{\overset{.}{R}^{2}}{R^{2}%
}-\frac{k}{R^{2}}\left(  1-k\frac{r^{2}}{R_{0}^{2}}+\cdots\right)
\ ,\label{36}\\
\ 8\pi P  &  \simeq-2\frac{\overset{..}{R}}{R}-\frac{\overset{.}{R}^{2}}%
{R^{2}}-\frac{k}{R^{2}}\left(  1-2k\frac{r^{2}}{R_{0}^{2}}+\cdots\right)  ,
\label{37}%
\end{align}
where $R(t)=R_{0}e^{g(t)/2}$. The crucial point in the above equations is that
$R(t)$ is still exact. Granted that the size of the local thermodynamic system
is sufficiently small with respect to the current radius, $r\ll R_{0},$ and
also the inequalities
\begin{equation}
\frac{\left\vert k\right\vert }{R^{2}}\ll\left(  \frac{\overset{.}{R}}%
{R}\right)  ^{2}\text{ and }\frac{\left\vert k\right\vert }{R^{2}}%
\ll\left\vert 2\frac{\overset{..}{R}}{R}+\frac{\overset{.}{R}^{2}}{R^{2}%
}\right\vert ,\text{ }k=\pm1, \label{38}%
\end{equation}
hold, the global pressure $P$ and the global density $\rho,$ approach to their
local expressions $P_{0}$ and $\rho_{0},$ hence the time dependent local field
equations become
\begin{align}
\ 8\pi P  &  \rightarrow8\pi P_{0}=-2\frac{\overset{..}{R}}{R}-\frac
{\overset{.}{R}^{2}}{R^{2}},\label{39}\\
\ 8\pi\rho &  \rightarrow\ 8\pi\rho_{0}=3\frac{\overset{.}{R}^{2}}{R^{2}}.
\label{40}%
\end{align}
Note that within the local thermodynamic system, the line element is%
\begin{equation}
ds^{2}\simeq dt^{2}-R(t)^{2}\left[  dr^{2}+r^{2}d^{2}\Omega\right]  ,
\label{41}%
\end{equation}
where $R(t)$ satisfies the inequalities in Equation (\ref{38}). Using the
local equation of state, $P_{0}=P_{0}(\rho_{0}),$ one can now determine
$R(t)$. Of course, after $R(t)$ is found, one has to check the inequalities in
Equation (\ref{38}) for self-consistency.

\section{The Field Equations and the Gibbs Energy}

In terms of the global pressure, $P(\rho,T),$ and the global density,
$\rho(\rho,T),$ the field equations [Eq. (\ref{4})] give two equations for the
four unknowns, $P,\rho,T,$ $R,$ as%

\begin{align}
8\pi P(\rho,T)  &  =-\frac{k}{R^{2}}-\frac{2\overset{..}{R}}{R}-\frac
{\overset{.}{R}^{2}}{R^{2}},\label{42}\\
8\pi\rho(P,T)  &  =\frac{3k}{R^{2}}+\frac{3\overset{.}{R}^{2}}{R^{2}}.
\label{43}%
\end{align}
Along with the definition of the gravitational temperature
\begin{equation}
T=\frac{\beta}{R},~\ \ \text{ }\beta=\left\{
\begin{tabular}
[c]{ccc}%
$\ 1/2\pi$ & $,$ & $k=1$\\
$0$ & $,$ & $k=0$\\
$i\sqrt{2}/\ 2\pi$ & $,$ & $k=-1$%
\end{tabular}
\ \ \ \ \ \ \ \ \ \ \ \ \ \ \ \ \ \ \right.  , \label{44}%
\end{equation}
we can complete this set with a \textit{global equation of state}, that is, a
relation between $P,\rho$ and $T,$ and solve for the four unknowns:
$\{P(t),\rho(t),T(t),R(t)\}$. However, the global equation of state requires
\textit{a priori} knowledge of the geometry of the universe, which is exactly
what we are trying to find out, hence this is not a viable option.

Instead, considering that the global variables $P$ and $\rho$ depend
implicitly on the local conditions described by $P_{0}$ and $\rho_{0},$ and
the gravitational temperature $T$, we rewrite the field equations [Eqs.
(\ref{42}) and (\ref{43})] as%
\begin{align}
8\pi P(P_{0},\rho_{0},T)  &  =-\frac{k}{R^{2}}-\frac{2\overset{..}{R}}%
{R}-\frac{\overset{.}{R}^{2}}{R^{2}},\label{45}\\
8\pi\rho(P_{0},\rho_{0},T)  &  =\frac{3k}{R^{2}}+\frac{3\overset{.}{R}^{2}%
}{R^{2}}. \label{46}%
\end{align}
We now have a system of two equations with six unknowns: $\left\{
P,\rho,P_{0},\rho_{0},T,R\right\}  .$ For a solution, we have to supplement
this set with four independent equations, which we take as

(i) The local equation of state $P_{0}=P_{0}(\rho_{0}),$ which is locally accessible.

(ii) The definition of gravitational temperature, which comes from QFT [Eq.
(\ref{44})].

(iii$-$iv) The two equations that the local pressure, $P_{0},$ and the local
density, $\rho_{0},$ satisfy [Eqs. (\ref{39}, \ref{40})]:
\begin{align}
8\pi P_{0}  &  =-\frac{2\overset{..}{R}}{R}-\frac{\overset{.}{R}^{2}}{R^{2}%
},\label{47}\\
8\pi\rho_{0}  &  =\frac{3\overset{.}{R}^{2}}{R^{2}}. \label{48}%
\end{align}

\subsection{Determining the Geometry of the Universe}

Using the definition of the gravitational temperature [Eq. (\ref{44})], and
the Equations (\ref{47}) and (\ref{48}), we can write the field Equations
(\ref{45}) and (\ref{46}) as%
\begin{align}
8\pi P  &  =-\frac{kT^{2}}{\beta^{2}}+8\pi P_{0}(\rho_{0}),\label{49}\\
8\pi\rho &  =\frac{3kT^{2}}{\beta^{2}}+8\pi\rho_{0}. \label{50}%
\end{align}
Eliminating $\rho_{0}$ among the above equations gives the Gibbs energy
density, $\rho_{(k)}(P,T),$ $k=0,\pm1,$ or the global equation of state,
$P_{(k)}(\rho,T),$ $k=0,\pm1,$ of the universe. Now the three geometries are
no longer isentropic, hence we can pick the $k$ value that has the lowest
Gibbs energy as the preferred geometry of the universe with respect to the
\textit{Friedmann thermodynamics}. Finally, with $k$ determined, using the
local equation of state, $P_{0}=P_{0}(\rho_{0}),$ we can solve Equations
(\ref{47}) and (\ref{48}) for $P_{0},$ $\rho_{0}$ and $R$ as functions of
\ time to complete the solution as
\begin{equation}
\left\{  P(t),\rho(t),P_{0}(t),\rho_{0}(t),T(t),R(t)\right\}  . \label{51}%
\end{equation}

\section{The Geometry of the Universe}

To demonstrate the above protocol for determining the geometry of the
universe, we consider the local equation of state $P_{0}=\alpha\rho_{0},$
$\alpha>0,$ which covers a wide range of physically realistic cases like
radiation for $\alpha=1/3$ and dust for very small values of $\alpha$. The
corresponding Gibbs energy densities, $\rho_{(k)}(P,T),$ $k=0,\pm1,$ can now
be found as%
\begin{align}
\rho_{(1)}(P,T)  &  =\left(  3+\frac{1}{\alpha}\right)  \frac{\pi T^{2}}%
{2}+\frac{P}{\alpha},\label{52}\\
\rho_{(0)}(P,T)  &  =\frac{P}{\alpha},\label{53}\\
\rho_{(-1)}(P,T)  &  =-\left(  3+\frac{1}{\alpha}\right)  \frac{\pi\left\vert
T\right\vert ^{2}}{4}+\frac{P}{\alpha}, \label{54}%
\end{align}
which for all $T$ and $P,$ indicate that the model with $k=-1$ has the lowest
Gibbs energy density. Hence with respect to the Friedmann thermodynamics and
the local equation of state, $P_{0}=\alpha\rho_{0},$ $\alpha>0,$ the stable
geometry of the universe is Lobachevskian. Of course, all these models being
locally equal, what determines their stability is their \textit{gravitational
entropy}. To complete the solution, we find $R(t)$ using Equations (\ref{47})
and (\ref{48}) as
\begin{equation}
R(t)=R_{0}\left[  \frac{3H_{0}(\alpha+1)}{2}(t-t_{0})+1\right]  ^{\dfrac
{2}{3(\alpha+1)}\ \ }, \label{55}%
\end{equation}
where $t_{0}$ is the current time (age) and $R_{0}$ is the current radius.

Finally, we check the inequalities in Equation (\ref{38}) for the
self-consistency of the model, which indicate that the current radius of the
universe has to be much larger than the visible universe:%
\begin{equation}
R_{0}\gg c/3\alpha H_{0}. \label{56}%
\end{equation}

Now the global equation of state is determined [Eq. (\ref{54})], the
consistency of the model can also be confirmed by the first approach by
supplementing the system of equations (\ref{42}-\ref{43}) with the global
equation of state [Eq. (\ref{54})]. It should be noted that our arguments
about the geometry are local and does not say anything about the global
topology of the universe, which is an other issue that the Friedmann
thermodynamics at this point can not answer. This prediction is consistent
with the WMAP data, which indicates that the universe may deviate from
flatness by as much as 1\%. Also, recently Liddle and Cortes [11] argued that
the observed large-scale asymmetry of the microwave background radiation may
be explained by assuming an open (Lobachevskian) universe slightly curved just
beyond the cosmic horizon.

\section{Conclusions}

The effective gravitational temperature we give is in terms of the
Maxwell-Boltzmann statistics and valid for regions large enough for the
effects of curvature to be important but yet small enough compared to the
current radius/size of the universe. We also showed that this method
reproduces the correct expressions for the Unruh temperature for the uniformly
accelerated reference frames and also the Hawking temperature for the black
holes [15]. However, for the accelerated observers and the black holes the
Green function expansions match to all orders, thus indicating that in
Friedmann thermodynamics the Maxwell-Boltzmann statistics is only
approximately true. In this regard, the gravitational temperature we give is
at best an effective temperature.

Using the concept of local thermodynamic equilibrium, we have suggested a
generalization of our gravitational temperature to arbitrary space-times [8].
For the Schwarzchild geometry, in the black hole limit, that is, when the
surface of the star approaches to its horizon, this temperature reduces
precisely to the Hawking temperature. This supports the view that black holes
are essentially the equilibrium state of any self gravitating system. In this
regard, the underlying statistical mechanics of the Friedmann thermodynamics
can not be the Maxwell-Boltzmann statistics. But considering that the black
hole thermodynamics obeys \ Maxwell-Boltzmann statistics, the attractor of the
Friedmann thermostatistics should be the Maxwell-Boltzmann statistics

An alternative to Maxwell-Boltzmann statistics is given by Tsallis et. al.
[17] and Tsallis [18], where the entropy is defined as%
\begin{equation}
S_{q}=k\left(  1-\sum_{i=1}^{W}p_{i}^{q}\right)  /(q-1). \label{57}%
\end{equation}
In the limit as $q\rightarrow1,$ the Tsallis entropy, $S_{q},$ reduces to the
Boltzmann entropy%
\begin{equation}
S=-k\sum_{i=1}^{W}p_{i}\ln p_{i}. \label{58}%
\end{equation}
With equal probability assumption, $S_{q}$ can be written as%
\begin{equation}
S_{q}=k\frac{\left(  W^{1-q}-1\right)  }{(1-q)}. \label{59}%
\end{equation}
When $S_{q}$ is maximized with the condition that the variance is finite and
the total probability is $1$, the probability of the $i$th state with the
energy $\epsilon_{i}$ being occupied becomes
\begin{equation}
p_{i}\varpropto\left[  1-(1-q)\frac{\epsilon_{i}}{kT}\right]  ^{1/(1-q)}.
\label{60}%
\end{equation}
In the limit as $q\rightarrow1,\ $this reduces to\textbf{ }the Boltzmann
weight factor
\begin{equation}
p_{i}\propto e^{-\epsilon_{i}/kT}. \label{61}%
\end{equation}

An important feature of the\textbf{ }Tsallis thermostatistics is that when two
independent systems with entropies $S_{q}(A)$ and $S_{q}(B)$ are combined, the
total entropy is given as%
\begin{equation}
S_{q}(A+B)=S_{q}(A)+S_{q}(B)+(1-q)S_{q}(A)S_{q}(B). \label{62}%
\end{equation}
This reduces to the sum of the individual entropies only when $q=1$. The
Tsallis thermodynamics is nonextensive, where $q$ is a measure of the
nonextensivity. For $q=1,$ the corresponding distribution function is a
Gaussian. For general $q<5/3,$ it is not possible to give an analytic
expression for the distribution function. However, from the central limit
theorem, the final distribution always tends to a Gaussian after many steps.
In this regard, Tsallis statistics is a possible candidate for the Friedmann thermodynamics.

In this exploratory but parameter free model, the left hand side of the
Einstein's field equations remain untouched, hence the vacuum field equations
are the same. However, even though the energy-momentum tensor is conserved,
geometry and matter are coupled through the \textit{global equation of state}
in a rather intricate way, that can not be covered by the conventional
approaches. It is this profound connection that we exploit in Friedmann
thermodynamics that allows us to predict the geometry of the universe [7-10].

\end{document}